\documentstyle[aps,prl]{revtex}

\tightenlines

\begin{document}
\draft
\title{Weak Field Hall Resistance and Effective Carrier Density Through
Metal-Insulator Transition in Si-MOS Structures}
\author{V.~M.~Pudalov$^{(a)}$, G.~Brunthaler$^{(b)}$, A.~Prinz$^{(b)}$,
G.~Bauer$^{(b)}$}
\address{$^{(a)} $P.\ N.\ Lebedev Physics Institute,\\
Moscow, Leninsky prosp. 53, Russia \\
$^({b})$ Institut f\"{u}r Halbleiterphysik,Johannes Kepler Universit\"{a}t,\\
Linz, A-4040, Austria}
\date{\today}
\maketitle

\begin{abstract}
We studied the weak field Hall voltage in 2D electron layers in Si-MOS
structures with different mobilities,
through the metal-insulator transition. In the vicinity of the
critical density on the metallic side of the transition, we have found
weak deviations (about 6-20 \%) of the Hall voltage from its classical
value. The
deviation does not correlate with the strong temperature dependence of the
diagonal resistivity $\rho_{xx}(T)$. The smallest deviation in $R_{xy}$
was found in the highest mobility sample exhibiting the largest variation
in the
diagonal resistivity $\rho_{xx}$ with temperature (by a factor of 5).
\end{abstract}


In the pioneering experiments by Fowler et al. \cite{fowler} performed on
Si-MOS structures in perpendicular magnetic field, the Shubnikov-de Haas
oscillations of the conductivity were found to be periodic
as a function of the gate voltage $V_g$.
This is consistent with the quasiclassical approach
\cite{ando}, in which the charge of the inversion layer $Q_{\rm inv}$ is
proportional to the applied voltage, as in a plain capacitor formed by the
metallic gate and the 2D layer. At low temperatures, when the bulk
conductance is frozen out and the charge in the
depletion layer does not vary
with gate voltage, variation of the capacitor charge with $V_g$ is related
to the inversion layer charge only:
\begin{equation}
Q_{{\rm inv}} = C (V_g - V_t).
\end{equation}
Here $C =dQ/dV_g$ is the geometric capacitance between the metallic gate and
the 2D carrier layer, $V_t$ is determined by the difference in work
functions of the Al- gate film and the 2D carrier layer, by the energy of
the bottom of the lowest subband in the confining potential well (zero-point
energy) and by the charge trapped in
depletion layer and at the interface
\cite{ando}. The charge in the
Si-MOS structure $Q_{{\rm inv}}$ was measured
directly \cite{semen} by integrating the current
entering (and quitting) the
structure during its charging (and discharging) with time.
The charge, $Q_{{\rm inv}}$, was found to be equal
(within 2\% uncertainty) to the charge of
the 2D carrier layer $Q_{{\rm 2D}} = e\times n_{{\rm ShdH}}$,
where $n_{{\rm ShdH}}$ is the density of carriers
participating in transport in the
Shubnikov-de\,Haas or QHE regime,
$n_{\rm ShdH} = (eB/h)\times i $, and
$i$ is the number of filled quantum
levels in a given magnetic field $B$.

The issue on
the carrier density was raised recently again in connection
with the metal-insulator (M-I) transition in 2D carrier systems at zero
magnetic field. The transition was found earlier to occur at a critical gate
voltage $V_{\rm gc}$ (where $V_{\rm gc} > V_t$) \cite{kravMIT}. By analogy
to the quasiclassical case \cite{ando},
and to the QHE-Insulator transitions
\cite{ES}, the critical gate voltage is interpreted
as a quantity corresponding
to a critical
carrier density $n_c = (dn/dV_g)(V_{{\rm gc}}- V_t)$. Recently, an
alternative interpretation was put forward \cite{dassarma} where the density
of carriers participating in zero-field transport was suggested to be equal
to $n_{{\rm eff}}= (dn/dV_g)V_g - n_c$, so that $n_{{\rm eff}} = 0$ at the
transition, i.e. at $V_g= V_{{\rm gc}}$.

The effective number of carriers is not a well defined
parameter close to the
M-I transition, and may,
apriori, be found different in different effects.
One of the most relevant effects capable to provide information on the
effective carrier density is the weak-field Hall effect where the Hall
resistance in the single-particle approximation \cite{smrcka} is inversly
proportional to the number of carriers:
\begin{equation}
R_{xy} \approx \frac{(\omega_c \tau)}{\sigma_0} [ 1-
\frac{1}{(\omega_c\tau_0)^2}
\frac{\Delta G(\epsilon_F)}{G_0}]
\end{equation}
Here $\sigma_0 =ne^2\tau /m^*$ is the diagonal conductivity at $B=0$,
$\omega_c = eB/m^*$ is the cyclotron frequency, and $\tau$,
the transport scattering
time at $B=0$. $G$ and $\Delta G$ are the monotonic and oscillatory parts of
the density of states. According to the theory \cite{altshuler},
electron-electron interaction affects $R_{xy}$ in the same order as
$\sigma_{xx}$. No corrections are expected in $\sigma_{xy}$ due to
interaction \cite{altshuler,houghton}, and, thus
$\delta R_{xy}/R_{xy} \approx 2 (\delta \rho_{xx}/ \rho_{xx}^0)$.

On the experimental side, the Hall resistance was measured
earlier across the
QHE-insulator transition \cite{krav91,pud93,kravhall}
and  found to remain close (within $\sim 10\% $) to its classical
value \cite{pud93} $R_{xy} = B/(ne)$ with $n$ given by Eq.~(1). Such
behavior of the Hall resistance is consistent with the
Hall insulator concept
\cite{lee}.
Low frequency ($\sim 3$\,Hz) ac-measurements of the capacitance "gate-2D
layer" \cite{krav91} have shown that, within a few \%, the capacitance 
remains unchanged across the QHE-insulator transition.  
With such a precision, the number of carriers, which is participating 
in low frequency charging-discharging processes, is described by the 
same Eq.~(1) in both, the insulating and metallic phases.  This 
observation also sets an upper estimate for possible variations of the 
trapped charge in the vicinity of the M-I transition: $(dV_t /dV_g) < 
10 \% $.
The contribution of the exchange and correlation
energy (negative compressibility) to the measured capacitance $C$ in Si-MOS
does not exceed $\sim 0.1\% $ \cite{nDOS}.

The above measurements of the Hall voltage and of the capacitance, however,
were done in {\em quantizing magnetic field}, which is known to change the
symmetry of the 2D system from orthogonal to unitary and thus may
disturb true zero-field properties of the metallic state. In this paper we
report measurements of the Hall voltage performed in such {\em weak magnetic
fields}, where Landau levels are not resolved, $\Delta G/G \ll 1$. We found
that the {\em weak-field Hall voltage remains finite through the
metal-insulator transition}. Deviation of the Hall voltage from the
quasiclassical value is within $(6-20)\% $ for different samples,
particularly at the critical gate voltage. In the assumption that the
inverse Hall voltage is a measure of the effective number of carriers,
$n_{\rm eff}$, we conclude the latter does not decrease to zero
at the M-I transition, i.e. at the critical gate voltage
$V_{\rm gc}$.

We studied three Si-MOS samples from different wafers: Si22 (with the peak
mobility, $\mu =33,000$\,cm$^2$/Vs at $T =0.3$\,K and the threshold voltage
$V_t \approx 0.4$\,V), Si4/32 ($\mu =8,000$\,cm$^2$/Vs, $V_t=2.1$\,V)
and Si-46 ($\mu= 1,350$\,cm$^2$/Vs, $V_t=6.5$\,V).
All samples had about the same
oxide thickness, $200 \pm 30$\,nm, and, correspondingly,
similar values of $dn/dV_g =
(1.05 \pm 0.15)\times 10^{11} $\,/Vcm$^{2}$.
The samples had potential probes lithographically defined
with accuracy of $1\mu$m. Geometric configuration of the
samples is shown in the inset to Fig.~1\,a,
where $w=800\,\mu$m and $l=1250\,\mu$m are the channel width and
intercontact distance, and $d = 25\,\mu$m is the width of the bulk diffusion
area contacting the 2D layer underneath the gate oxide. For Hall voltage
measurements at low density where the contacts and the sample area have high
resistance, we used
battery operated electrometric amplifiers with input
current $< 10^{-14}$\,A. Four-probe measurements were taken by ac lock-in
technique at (3 - 7)\,Hz, and partly by dc-technique.

Figure~1\,a
demonstrates that minima in the Shubnikov-de\,Haas oscillations in $\rho_{xx}$
(which correspond to the minima in the density of states between four-fold
degenerate Landau levels) are equidistant in the gate voltage scale. The
carrier density calculated from the Shubnikov-de Haas minima was found to be
independent of magnetic field (in the range 0.5 to 5 Tesla) and of temperature
(for 0.3 to 1.4\,K), within uncertainty of (1-2)\%. For Hall
voltage measurements, we have chosen magnetic
fields of
0.2 and 0.3\,Tesla, for
high and low mobility samples, respectively.
This field is large enough
to suppress the quantum interference corrections
to conductivity \cite{JETPL97b}.
On the other hand, this field is low enough to
keep the oscillatory part in the density of states
small, for all gate voltages and
temperatures down to 300\,mK. Figure~1\,b shows
the density dependence of $\omega_c \tau $ calculated
from the measured mobility, for the sample Si22
in the field of 0.2\,T. For two other samples,
the $\omega_c \tau $ value is
evidently smaller. Due to the
smallness of $\omega_c\tau \leq 1$,
and of  $\Delta G/G \ll 1$,
the second term in Eq.~(2) is negligibly small.

The samples Si22 and Si4/32 exhibited well pronounced exponential decay in
resistivity for $V_g > V_{{\rm gc}}$ as temperature decreases, and,
correspondingly, a symmetrical scaling plot \cite{kravMIT,mautern}. The most
disordered sample Si46 displayed an activated $\rho(T)$-dependence in the
insulating state (for $V_g < V_{{\rm gc}}$) and a weak, almost linear,
``metallic'' temperature dependence for $V_g >V_{{\rm gc}}$ \cite{mautern}.
The overall density dependence of the resistivity for Si46 at 6 different
temperatures,
shown in the inset to Fig.~2,
demonstrates a critical behavior around $V_{\rm gc} =14.4$\,V.
Figure~2 shows the effective
"Hall-density" $n_{{\rm Hall}} = B/(eR_{xy})$ calculated from $R_{xy}$
measured at $T=0.29$K. The dashed line depicts $n_{{\rm ShdH}}$ vs gate
voltage, calculated from the period of the Shubnikov-de Haas effect. As $V_g$
decreases, the effective Hall density slightly deviates from the classical
linear dependence, and then falls quickly to zero, deep in the insulating
state at $V_g = 12$\,V. Just at the critical gate voltage $V_{\rm gc}=14.4$\,V,
the Hall-density is by 5\% {\em larger} than the classical value given by
Eq.~(1).

For samples with higher mobility, the critical carrier density $n_c$ is
lower and the critical resistivity $\rho_c$ is higher \cite{mautern}. By
this reason, the admixture of the longitudinal voltage produces large
distortions (oscillating with gate voltage) of the measured Hall voltage. As
Fig.~3\,a shows, this admixture can be reduced by an order of magnitude by
subtracting the results taken
for opposite magnetic field directions.
The Hall resistance for sample Si4/32 at low density is significantly
larger than the quasiclassical value $B/(en_{{\rm ShdH}})$. The deviation in
the effective Hall-density,
$\delta n_{{\rm Hall}}= n_{{\rm Hall}}-n_{\rm ShdH}$,
calculated from the measured $R_{xy}$ values for six temperatures is
plotted in Fig.~3\,b. At high density and high temperature, the deviation in
the Hall density tends to zero. As temperature decreases to 0.3\,K, the
deviation raises to almost 20\%, and seems to saturate. Oscillations and
scattering in the Hall density data (seen in the range of low $V_g$)
are due to residual admixture of the
large longitudinal $V_{xx}$ voltage into the small Hall voltage.

Finally, as shown in Fig.~4, the deviation in the Hall density for the high
mobility sample Si22 depends on gate voltage and temperature
qualitatively similar but 3 times weaker than for Si4/32. As
gate voltage increases, the disagreement between $n_{\rm Hall}$
and $n_{\rm ShdH}$ becomes less than the measurement uncertainty.
We must note, however, that the absolute value of $n_{\rm ShdH}$
has the uncertainty about
(1-2)\% for Si22 and 4\% for Si4/32.
By this reason, the true position
of the ``zero'' on the vertical scales in Fig.~4 and
Fig.~3\,b
is defined within the uncertainty of
 $(1-4)$\%, correspondingly.
Although the deviation of the Hall density in the vicinity of $V_{\rm gc}$
is small for all samples, $\delta n_{\rm Hall}/n_{ShdH} \ll 1$,  it
is much larger than the error bars.
Due to the charge neutrality in the total
Si-MOS structure, Eq.~(1), the nonzero value of the $\delta n_{\rm Hall}$
indicates either a lack of the
Drude-Boltzmann interpretaion of the Hall voltage
in the vicinity of $V_{\rm gc}$, or
a noticable contribution of carriers exchange between the 2D layer
and the shallow potential traps.

In the framework of the Drude-Boltzmann model,
the effective carrier density $n_{\rm Hall}$ for all samples remains
close to the classical value
through the metallic range of densities $V_g >V_{\rm gc}$,
where the resistivity $\rho_{xx}$ strongly varies  with temperature.
We conclude, in the same
framework, that the strong exponential
drop in $\rho _{xx}(T)$ as $T$ decreases \cite{kravMIT},
is associated with an anomaly in
the scattering time or in the transport mechanism, rather than with
carrier density.
The insets to figures~3\,a and 4 show that the variations
of the diagonal resistivity with temperature
(by a factor of 2.5 and of 5.5
for Si4/32 and Si22, correspondingly)
are much larger than that in $R_{xy}$, over the same range
of density and of temperature.
The lack of a linear relationship between
$\delta R_{xy}$ and $\delta \rho_{xx}$
in the vicinity of $V_{\rm gc}$,
indicates that at least one of these two quantities
is not related to the interaction quantum corrections
\cite{altshuler} (with the reservation that the
theory may be not valid for the strong interaction case,
$r_{s}\sim 3-10$).

In conclusion,
we have measured the weak field Hall resistance in $n-$Si-MOS samples
through the metal-insulator transition.
We found no strong changes in $R_{xy}$
(comparable to those in $\rho_{xx}(T)$)
and found no signatures
of a complete carriers freeze-out at $V_{g}=V_{\rm gc}$.
However, for low
density and low temperatures,  the
Hall voltage in different samples was found to deviate from the classical value
by about  (6-20)\%.
The deviation in $R_{xy}$ does not correlate with the
strong temperature dependence of the diagonal resistivity $\rho _{xx}(T)$.
Particularly, among the samples studied,
the smallest value of the deviation in $R_{\rm xy}$ (by 6\%)
was measured in the high mobility sample Si22, where
the diagonal resistivity $\rho _{xx}$ varies
most strongly (by 5.5 times) in the same temperature range.

V.P.  acknowledges help by M.\ D'Iorio and E.\ M.\ Goliamina for 
sample processing, and discussions with B.\ Altshuler and D.\ Maslov. 
During completion of our manuscript, we learned that
measurements of the Hall resistance in Si-MOS
across the transition were done also by  D.\ Simonian, K.\,M.\ Mertes, 
M.\,P.\ Sarachik, and S.\,V.\ Kravchenko. 
The work was supported by RFBR, by the Programs ``Physics of 
solid-state nanostructures'' and ``Statistical physics'', by INTAS, 
NWO, and by FWF P13439, \"{O}NB 6333 and GME, Austria.

\begin{figure}[tbp]
\caption{a) Shubnikov-de Haas oscillations in $\rho_{xx}$ measured on the
sample Si22 vs gate voltage at $T=0.29$\,K and $B=2$\,T. Dashed line and
full dots demonstrate a linear dependence between the number of the quantum
level and the gate voltage, from which the $n_{{\rm ShdH}}$ density is
calculated. Inset shows the sample geometry.
b) $\omega_c \tau $ vs density
at field $B=0.2$\,T.}
\end{figure}

\begin{figure}[tbp]
\caption{Resistivity at zero magnetic field (left Y-axis)
vs gate voltage  for the sample Si46 for 7
temperature values. Hall density
(right Y-axis) as a function of gate voltage
at  $B=0.3$\,T and $T =0.29$\,K.
Dotteded line depicts $V_{\rm gc}$,
dashed line is for the density $n_{\rm ShdH}$
calculated from Shubnikov-de Haas oscillations.}
\end{figure}

\begin{figure}[tbp]
\caption{a) Hall resistance as a function of the density, measured at two
opposite field directions with sample Si4/32 at $T=2$\,K and $B=0.2$\,T.
Dotted curve represents an averaged resistance, $<R_{xy}>=(R_{xy}(+B)
-R_{xy}(-B))/2$. Bold dashed curve - classical dependence $B/ne$. b)
Deviation in the ``Hall-density'',
$\delta n_{\rm Hall} =n_{\rm Hall} -n_{\rm ShdH}$, measured
at $B= \pm 0.2$\,T for six temperatures. Vertical
dashed lines mark the critical density $n_c$.
The inset shows $\rho_{xx}$ vs temperature for
11 densities, 2.4, 2.5, 2.6, 2.7, 2.8, 3, 3.2, 3.4, 3.7, 4.7, 5.7,
in unites of $10^{11}$cm$^{-2}$.}
\end{figure}

\begin{figure}[tbp]
\caption{Deviation in the ``Hall-density'',
$\delta n_{{\rm Hall}} =n_{{\rm Hall}} -n_{{\rm ShdH}}$,
measured at $B= \pm 0.2$\,T with sample Si22 for four
temperatures. Error bars indicate typical uncertainty for
different temperatures and densities.
Dashed line is a guide to the  eye.
Dotted vertical line marks the critical density $n_c$.
The inset shows $\rho_{xx}$ vs temperature for 11 densities,
1.5, 1.7, 1.8, 1.9, 2.1, 2.4, 2.9, 3.34, 4.3, 5.5, 7.9,
in unites of $10^{11}$cm$^{-2}$.}
\end{figure}


\end{document}